# Experimental and micromagnetic investigation of texture influence on magnetic properties of anisotropic Co/Co$_3$O$_4$ exchange-bias composites

Kurichenko Vladislav L.[1*], Karpenkov Dmitriy Yu.[1,2], Degtyarenko Alena Yu.[3]

[1] National University of Science and Technology "MISIS", Leninskiy prospect, 4, 119049 Moscow, Russia
[2] Lomonosov Moscow State University, Leninskie Gory 1, 119991 Moscow, Russia
[3] Ginzburg Center for High Temperature Superconductivity and Quantum Materials, P.N. Lebedev Physical Institute of the RAS, 119991, Moscow, Russia
[*] Corresponding author. E-mail: vkurichenko@misis.ru

**Abstract**

The exchange interaction between nanostructured components is an effective way to enhance the magnetic properties of materials. This effect is used in exchange-coupled magnetic composites, which properties are governed by synergetic impact of constituent phases. Texturing is one of the problems that needs to be solved in order for such composites to be used in industry. In this work, we performed experimental and micromagnetic investigation of the exchange-bias properties in Co/Co$_3$O$_4$ nanocomposites based on nanorods array. Specifically, we investigated how the resulting properties will change depending on the nanorods texture. Our experiments proved previous theoretical calculations of exchange-bias nanorods array-based composites that showed that magnetic properties of such materials are dependent on the internal texture. Tailoring the texture can lead to either increase of exchange-bias field, or to enhancement of material coercivity.

Keywords: micromagnetic modelling, exchange-coupled composites, nanorods, exchange-bias composites, texture

## 1. Introduction

In 1991, Kneller and Hawig proposed a new type of magnetic material, called 'exchange-coupled', in which the magnetization of the hard magnetic phase can be increased by adding a magnetically soft one with a high saturation magnetization ($M_S$) value. However, it is necessary that the phases are exchange-coupled at the nanoscale level at the interface [1]. Such 'exchange-coupled' composites are considered as one of the most effective ways to enhance the magnetic properties of materials.

Exchange-coupled somposites are divided in two groups. The first one is 'exchange-spring composite', which is based on soft/hard or hard/hard ferromagnetic materials (FM/FM). In such composite magnetically hard phase is responsible for enhanced anisotropy, while magnetically soft phase provides high magnetization [2]–[4]. The other type is 'exchange-bias composite', based on ferromagnetic and antiferromagnetic materials (FM/AFM). In such composite when the phases are coupled and are field-cooled below Neel temperature of an antiferromagnetic phase, hysteresis loop shift should be observed [5]–[8].

Currently, in order to use exchange-coupled composites for the production of permanent magnets, it is necessary to find a solution to the problems that arise at various stages of obtaining such permanent magnets: texturing, compacting and scaling up production [9]. The last two problems are out of scope of this article, so we will focus on the texturing.

The lack of preferred crystallographic orientation and grain size homogeneity in obtained exchange-coupled nanocomposites leads to deterioration of the magnetic properties of such composites. E.g. Li et al [10] have shown the importance of texturing on resulting magnetic properties of SmCo/FeCo composites, as texturing enhanced energy product by 50 % as compared to the isotropic composite. In another work it was shown that aligning of Co/CoO nanorods array can lead to an increased exchange-bias field, which was almost two times higher than for isotropic samples [11].

In their theoretical works, Patsopoulos et al investigated the physics behind crystallographic orientation of grains and exchange-bias properties by Monte-Carlo simulations of nanorods with a ferromagnetic core and an antiferromagnetic shell (Co/CoO). The authors showed that with a polycrystalline CoO shell but a single-crystal Co core, an increase in the coercivity should be observed, accompanied with a decrease in the effect of the exchange coupling. On the contrary, for a single-crystal shell an increase in the effect of the exchange-bias together with a decrease in the coercivity are observed [12], [13].

Thus, by creating the internal texture in the nanocomposite, one can not only enhance but also manipulate the resulting magnetic properties of the samples, in terms of coercivity values and exchange-bias effect.

The influence of texture on Co nanorods magnetic properties was studied in the literature[14]–[17]. However, the influence of the texture of Co/Co oxide composite on resulting exchange-bias properties has not yet been studied in the literature experimentally, so it was the aim of this work. Cobalt was chosen as a material of interest, since its oxides are antiferromagnetic, hence it is possible to obtain the FM/AFM nanocomposite by oxidizing the sample [18]. Nanorods array structure was investigated, as it is initially anisotropic, which can allow to further increase the magnetic properties of the material [19].

## 2. Methods

Co nanorods were synthesized via electrochemical deposition into polycarbonate membranes with pore diameter of 100 nm [20]. Details of the electrodeposition are presented in Supplementary Information. By changing the parameters of electrodeposition, it is possible to obtain nanorods of cobalt with different texture [21]. Therefore, two different current densities were selected: 2 and 4 mA/cm$^2$.

The technique for obtaining exchange-bias composites based on cobalt nanorods consists in the oxidation of its surface, since it is known that cobalt oxides are antiferromagnetic. Due to usage of polycarbonate membranes, which have glass transition temperature of 150 °C, the oxidation treatment temperature was limited by 140 °C, the duration of the treatment under air atmosphere was 12 h. Even though this temperature is quite low for oxidation, it could still lead to production of 4 nm-thick oxide layer according to the article describing kinetics of cobalt oxidation [22].

Micromagnetic calculations of the exchange-bias nanocomposites were performed in Mumax3 [23]. The model is described in Supplementary Information.

Morphology of the samples was investigated using scanning electron microscope (SEM) TESCAN Vega 3 SB with Energy dispersive X-ray spectroscopy (EDX) setup. Crystal structure and phase composition were studied by X-ray diffraction method (XRD) on Difrei 401 under Cr radiation. Magnetic mesurements at low temperatures were performed using the Quantum Design PPMS-9 system. Magnetic measurements at room temperatures were performed on Lake Shore 7407.

## 3 Results and discussion

At the first stage, nanorods with different texture were obtained. Figure S1 shows potential curves during deposition. Results of the XRD analysis of nanorods are presented on Figure 1a. Diffraction peaks from copper substrate are revealed on the pattern. For both samples, the cobalt phase posesses a *hcp* structure. However, during deposition with lower current density, nanorods possesses preferred texture in the (002) direction. But when higher current density is used, the texture in the samples changes, and additional peaks from (100) and (101) planes can be observed. No peaks corresponding to oxide phases were revealed. Therefore, the sample deposited at lower and higher current density will be called in the article as 'textured' and 'non-textured', respectively.

Figure 1b depicts the SEM images in back-scattered electrons (BSE) signal of the nanorods after membrane dissolution. The obtained micrographs show that the nanorods have a uniform length of about 4.8 μm and a diameter of 120 to 140 nm.

Hysteresis loops for samples are shown on Figure 2. The magnetic field was applied parallel and perpendicular to the nanorods axis, the corresponding field direction are marked as ∥ and ⊥. In the first case (Figure 2a) the field direction coincides with easy axes of both the magnetocrystalline and shape anisotropies, but for the hard axis of the nanorods array as a whole, which can be considered as a thin film consisting of free-standing nanorods. Thus, when the field is applied perpendicular to the nanorods axis (Figure 2b), it is applied along the film surface. The obtained loops correspond to the arrays of the separated nanorods, which are located at a sufficient distance from each other to suppress the interaction between them. Hence, the hysteresis loops observed when the field was applied perpendicular to the nanorod axis correspond to the hard axis despite the fact that sample was magnetized along its plane.

To explain such behaviour we can use proof by contradiction method while comparing the values of shape anisotropy constant $K_d$ for the whole array and the magnetocrystalline constant $K_1$ of the individual nanorod. The former can be estemated as follows [24]:

$$K_d = \frac{1}{2} 4\pi N M_s^2 \qquad (1)$$

where $N$ is the demagnetizing factor and $M_s$ is saturation magnetization. As a starting point, we assumed that we had solid dense film, consisting of fully-interacted nanorods.

Hence, in the case when the magnetic fiels is applied along the nanorod axis and perpendicular of the film plane, we should use the value of $N = 4\pi$. Saturation magnetization of cobalt is 1400 Gs [25]. This gives us the value of $K_d = 12 \cdot 10^6$ erg/cm$^3$. $K_1$ of *hcp* cobalt is $5 \cdot 10^6$ erg/cm$^3$ [26]. Those values are comparable. It contradicts with the experimental results, because there is almost no impact of the shape anisotropy of the sample, the obtained field dependences correspond to easy axis measurement. So, we can say that in our sample $N$ is not $4\pi$, meaning that the nanorods in the array do not interact with each other fully.

The loops also show that nanorod texturing makes it possible to increase the coercivity from 500 Oe to 1 kOe (for non-textured and textured sample, respectively). There is also an increase in $M_R/M_S$ ratio during texturing from 0.29 to 0.47.

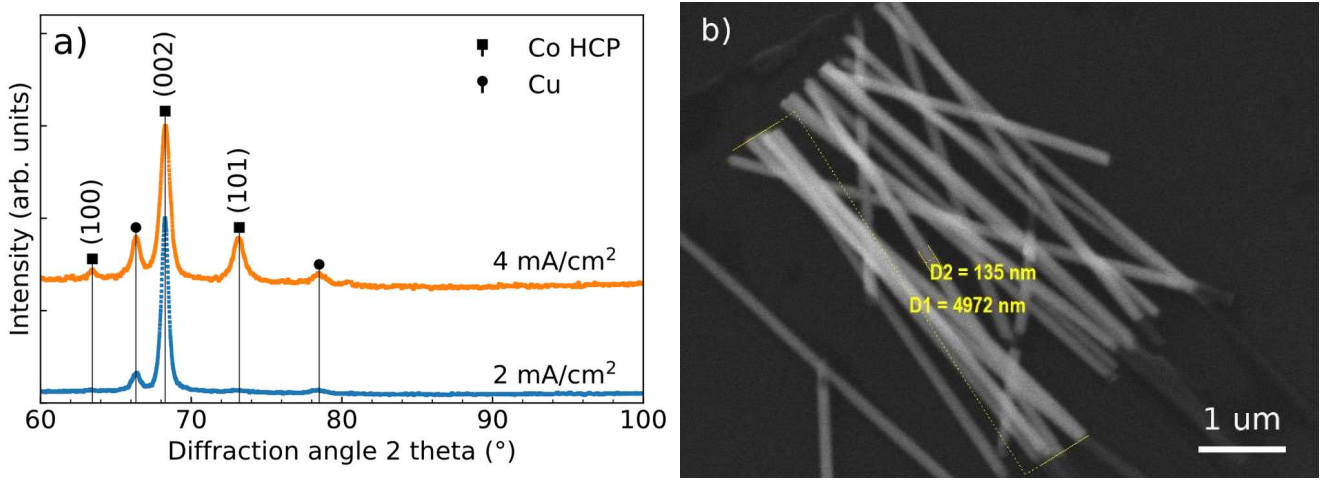

**Figure 1.** a) Diffraction patterns of nanorods deposited with different current densities. b) SEM microphotographs

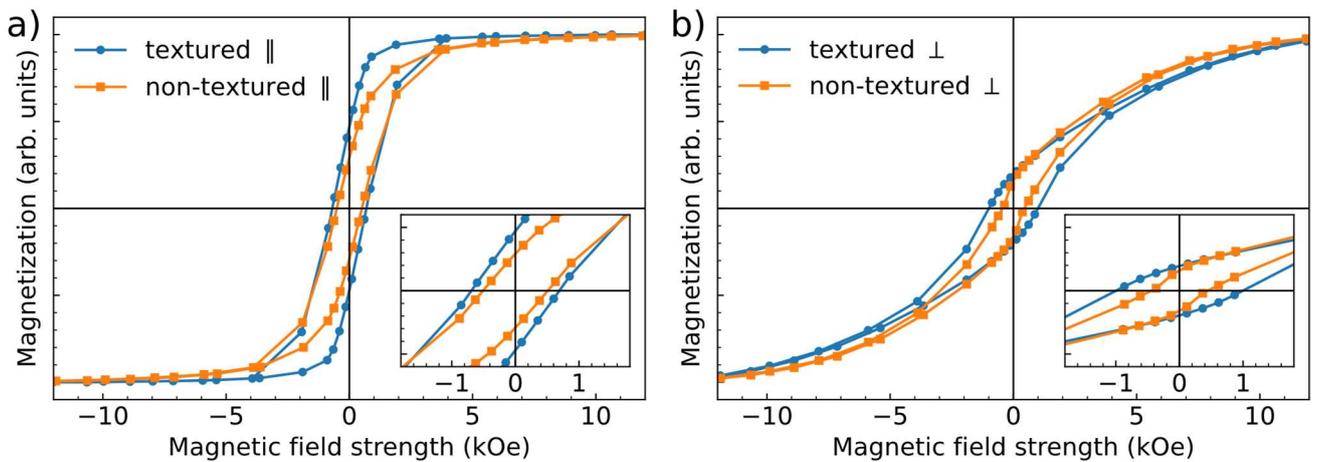

**Figure 2.** Hysteresis loops of nanorods with different texture. a) The field was applied parallel to nanorods axis. b) The field was applied perpendicular to nanorods axis.

In order to further investigate interactions in nanorods array, two other methods of characterization were used: Henkel plots [27] or Kelly plots [28] and First-Order-Reversal-Curves (FORC) analysis [29].

The Henkel plots are obtained by measuring the sample in a certain procedure, which is described elsewhere [27]. After the measurements, $\delta m(H)$ devendence can be plotted. Magnetostatic interaction between particles leads to negative $\delta m(H) < 0$. If the particles are exchange-coupled, it results in mainly positive $\delta m(H)$ [30]. The higher intensity of the peak would mean the higher corresponding interaction [31], [32].

Figure S2 shows corresponding Isothermal remanent magnetization (IRM) and DC demagnetization remanence (DCD) curves measured along the nanorod axis. $\delta M(H)$ dependence (Kelly plots) was calculated from the obtained curves and is presented on Figure 3a (inset shows corresponding Henkel plots). According to the obtained Henkel plot in both samples magnetostatic interactions between nanorods surpass the exchange one, since the $m_{dcd}(m_{irm})$ curves deviate from the dashed line, which shows the case of no interactions [32].

$\delta M(H)$ (Kelly plot) dependence shows that for the textured sample the dipolar interactions are higher than for the non-textured sample. However, the Kelly plots can be only considered as a qualitative analysis of interactions in the sample and both magnetostatic and exchange interactions have an effect on obtained curve [33]. Moreover, if there is a texture in easy axes distribution, the magnetostatic interaction can differ depending on orientation of those axes [30]. So in our case non-textured sample possesses reduced dipolar contribution, which can be attributed to lower remanence magnetisation.

The FORC analysis was also performed for the samples to quantatively study the interaction in the samples. Figure S3 shows corresponding hysteresis loops that were obtained in order to calculate FORC. Figure 3b shows the FORCs corresponding to both samples. It is seen that both samples have one clear peak, which is a bit more extended in $H_U$ axis in the case of the textured sample. In order to get more information about the FORC distributions, the maximums of the FORCs were estimated, which are presented as white dashed lines on figures. After that, $\rho(H_C)$ (Figure 3c) and $\rho(H_U)$ (Figure 3d) dependences corresponding to the white dashed lines were plotted for both samples.

The $\rho(H_C)$ dependence presented on Figure 3c can be described as a coercivity distribution profile. Coercivity, in turn, depends on the microstructure of the sample. This can explain the fact that for the textured sample there is one peak of coercivity observed, but for the non-textured one some distribution in coercivities occured. It can be explained by the miorientation of anisotropy axes directions of the different grains in the non-textured sample.

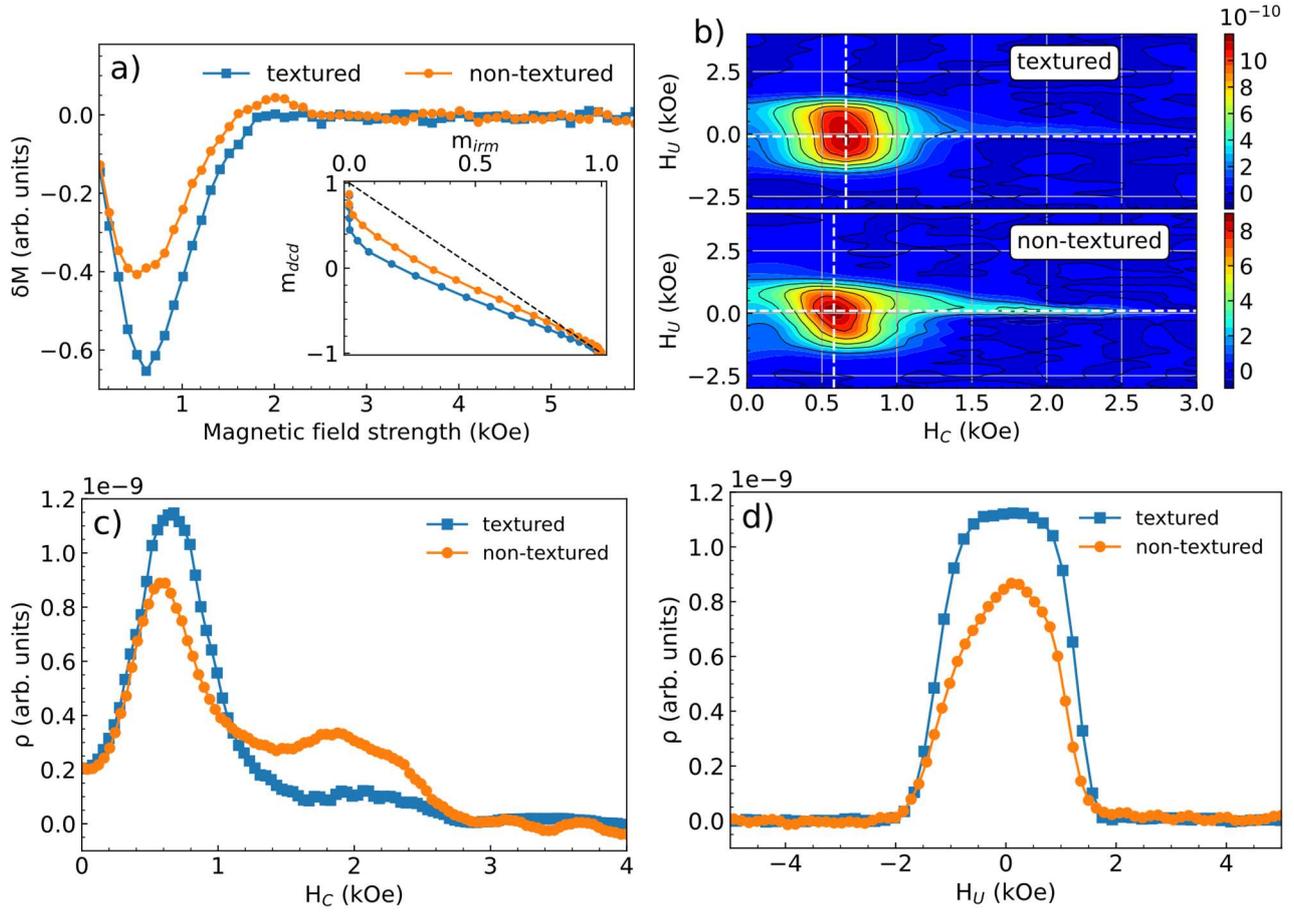

**Figure 3.** a) δm(H) dependences for the samples. b) FORC of the samples. c) ρ($H_C$) dependences. d) ρ($H_U$) dependences.

In turn, ρ($H_U$) dependence presented on Figure 3d can be described as an interaction field distribution. It is seen that for the textured sample corresponding ρ peak values are higher and are more flat. Due to the size effects influence of interaction field distribution, explored in [34], the interaction field peak can be broadened either due to increased number of interacting nanorods and their diameter or decreased length and interwire distance. Since we are using the same membranes for nanorods production, those effects are not the reason of the increased interaction mean field. Thus, this behaviour can be attributed to the fact that in the textured sample the effective the stray fields should be more extended, since individual grains' anisotropy axes are pointing in the same direction, which leads to increased net magnetisation. As a consequence, increase in the mean interaction field between nanorods is observed. In the non-textured sample the stray fields are less pronounced, due to grains' easy axes misorientation, thus decreasing magnetostatic interaction between the nanorods. These conclusions are consistent with the fact that $M_R/M_S$ ratio in the textured sample is higher.

After investigating magnetic properties of nanorods array and validating texture existence, we should turn to exchange-bias composites production. In case of cobalt, its surface oxidation is possible after synthesis, leading to the formation of antiferromagnetic oxide phases. However, even though transmission electron microscopy (TEM) photos showed presence of thin $Co_3O_4$ oxide layer on sufrace of as-prepared samples (Figure S4), magnetic measurements at low temperature did not show any expected loop shift when measured by zero-field-cooled (ZFC) and field-cooled (FC) protocols (Figure S5).

Therefore, the samples were additionally oxidized. TEM photo of the textured sample after oxidation is presented on Figure S6. It is seen that thickness of the oxide layer has increased. Electron diffraction pattern shows that the sample has $Co_3O_4$ phase. The resulting hysteresis loops are presented on Figure 4. The results show that in the textured sample there is a loop shift from 1.57 to 1.75 kOe, which is associated with the effect of the exchange coupling in the sample. For an untextured, this shift is also observed, but it is less pronounced, as compared with the overall increase of coercivity.

In order to estimate the effect of exchange-bias the corresponding exchange field $H_{EB}$ can be calculated using the equation (2):

$$H_{EB} = \frac{(H_{C1}+H_{C2})}{2} \quad (2)$$

where $H_{C1}$ is the coercivity of the ascending magnetization curve, $H_{C2}$ – coercivity of the descending magnetization curve. The corresponding values of exchange field $H_{EB}$ for textured and non-textured samples are -60 and -35 Oe, respectively. This confirms the conclusions made in one of the works devoted to the modeling of exchange-coupled composites, according to which the effect of exchange-bias should increase during texturing of samples and in non-textured samples should lead to the increase of coercivity and decreased exchange-bias [13].

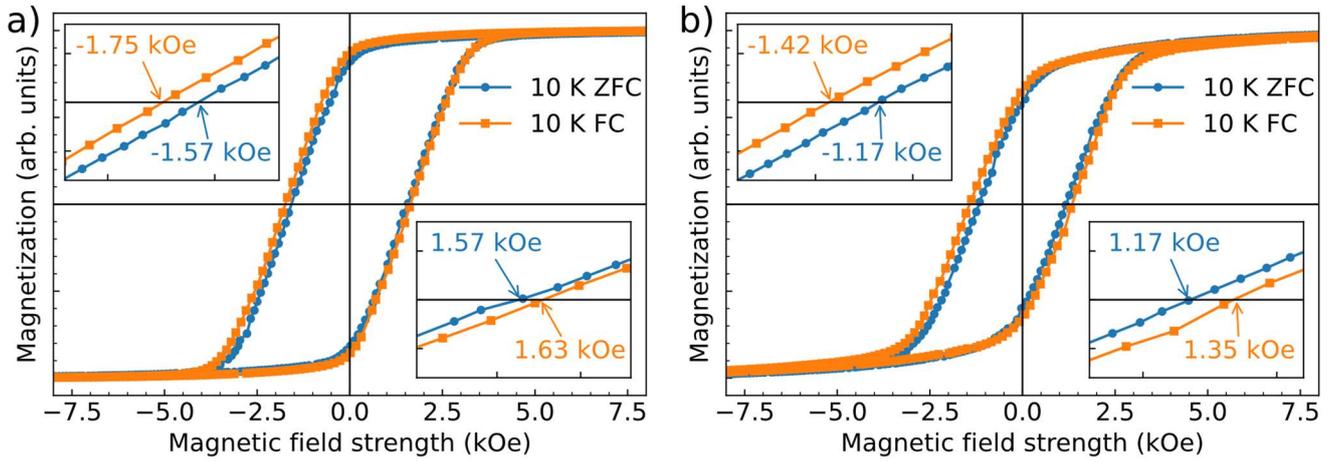

**Figure 4.** Hysteresis loops for the oxidized samples. a) textured sample. b) non-textured sample.

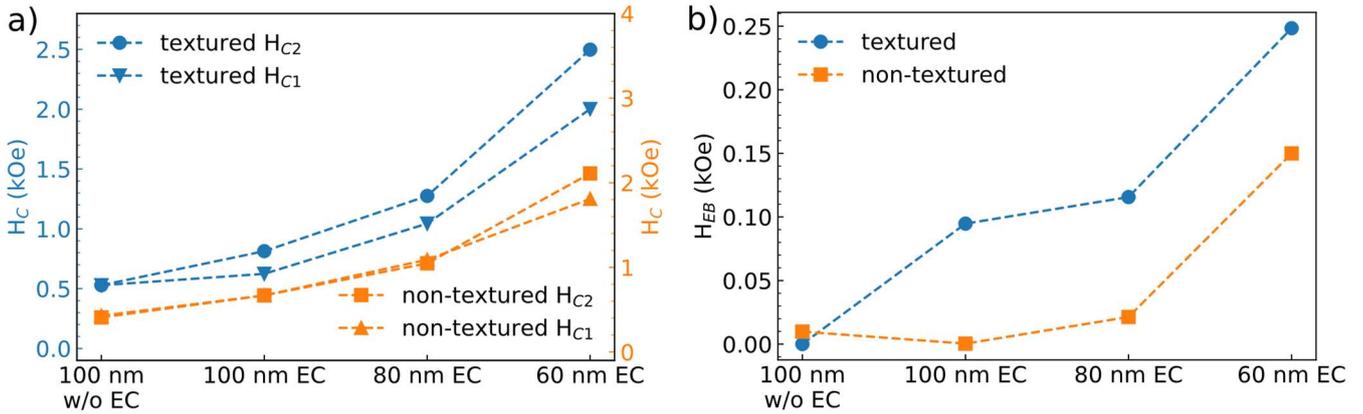

**Figure 5.** a) $H_{C1}$ and $H_{C2}$ values extracted from calculated hysteresis loops. b) $H_{EB}$ values extracted from calculated hysteresis loops.

Since the experimentally obtained results revealed that the influence of the crystallographic texture on the exchange-bias properties is not significant, micromagnetic simulations in Mumax3 were performed. Figure 5a shows calculated dependences of $H_{C1}$ and $H_{C2}$ for non exchange-coupled (EC) nanorod with 100 nm diameter, as well as for exchange-coupled nanorods with lower diameters. It is seen that for both samples $H_{C1}$ equals to $H_{C2}$ when there is no exchange coupling between FM/AFM phases in the sample. For the non-textured sample this behaviour maintains for nanorods when their diameter is more than 60 nm even when exchange-coupling is present. However, for smaller-diameter nanorods an exchange-bias behaviour presents, since $H_{C1}$ is not equal to $H_{C2}$, which is consistent with previous calculations [12]. Figure 5a shows $H_{EB}$ values calculated using Equation (2). For the non-EC sample $H_{EB}$ equals to zero. In turn, for textured 100-nm-thick exchange-coupled nanorod $H_{EB}$ value is enhanced, as compared with the value for the non-textured sample. Both calculations agree with experimental results presented on Figure 4. Moreover, it is seen that when diameter of the ferromagnetic core decreases, the value of $H_{EB}$ increases for both samples, but it is higher for the textured nanorods. It can be explained by the fact that the exchange-bias effect in FM/AFM composite inversely depends on the FM phase thickness [8].

## Conclusions

To conclude, in this article exchange-bias nanorods array-based $Co/Co_3O_4$ composites by means of electrodeposition in polycarbonate membranes were synthesized. After obtaining Co nanorods with different texture they were additionally oxidized in order to form $Co/Co_3O_4$ exchange-bias composites. Experimental results show that in the textured sample there is an exchange-bias field of - 60 Oe. In the non-textured sample this value was found to be - 35 Oe. The simulation results obtained by micromagnetic calculation in Mumax3 showed the same dependence of exchange-bias field when changing the texture of the sample. Moreover, it was shown that when decreasing the thickness of ferromagnetic phase, the exchange-bias field increases. Therefore, by texturing the sample it is indeed possible to adjust magnetic properties of exchange-bias composites. In order to achieve more exchange-bias field one should obtain textured sample with decreased ferromagnetic phase dimensions.

## Acknowledgements

The reported study was funded by RFBR according to the research project № 20-33-90154. The study of the structure was carried out on the equipment of the Center Collective Use «Materials Science and Metallurgy» in NUST MISiS. Magnetic experiments at helium temperature were obtained using equipment of the Lebedev Physical Institute's Shared

Facility Center. We acknowledge Luchnikov L.O., Muratov D.S. and Komlev A.S. for the help with methodology and investigaton.

**CRediT author statement**

**Kurichenko V.L.**: Conceptualization, Methodology, Formal analysis, Investigation, Data Curation, Writing - Original Draft, Writing - Review & Editing, Visualization. **Karpenkov D.Yu.**: Formal analysis, Investigation, Writing - Review & Editing, Supervision. **Degtyarenko A.Yu.**: Investigation.

# Supplementary information

# Experimental and micromagnetic investigation of texture influence on magnetic properties of anisotropic Co/Co$_3$O$_4$ exchange-bias composites

Kurichenko Vladislav L.[1*], Karpenkov Dmitriy Yu.[1,2], Degtyarenko Alena Yu.[3]

[1] National University of Science and Technology "MISIS", Leninskiy prospect, 4, 119049 Moscow, Russia
[2] Lomonosov Moscow State University, Leninskie Gory 1, 119991 Moscow, Russia
[3] Ginzburg Center for High Temperature Superconductivity and Quantum Materials, P.N. Lebedev Physical Institute of the RAS, 119991, Moscow, Russia
[*] Corresponding author. E-mail: vkurichenko@misis.ru

## Methods

*Nanorods electrodeposition technique*

Isopore membranes made by Merck Millipore with pore diameter of 0.1 μm and membrane thickness of 20–25 μm were used. A thin layer of copper was deposited on one side of the membrane to make electrical contact. The copper-coated membrane acts as a working electrode in a two-electrode circuit and is placed opposite the graphite counter electrode. The composition of the electrolyte for the deposition of cobalt nanorods: CoSO4·7H$_2$O - 84.33 g/L, H$_3$BO$_3$ - 45 g/l, pH = 6 was adjusted with NaOH solution. Two different current densities were selected: 2 and 4 mA/cm$^2$. Deposition time was 30 and 15 minutes, respectively.

*Mumax3 micromagnetic calculations details*

The model that was used is similar to the one that was described in our previous article [1]. As proposed in one of the articles, AFM phase can be implemented in Mumax3 as an FM phase with some effective parameters, which are summarized in Table S1. Saturation magnetization of cobalt oxide was the same as for cobalt phase, so that sufrace energy density if equally divided between cells [2].

Table S1 – Materials parameters for micromagnetic calculations (taken from [2])

| Parameters | $M_S$ (kA/m) | $A_{ex}$ (pJ/m) | $K_1$ (kJ/m$^3$) |
| --- | --- | --- | --- |
| Cobalt | 1400 | 30 | 20 |
| Cobalt oxide | 1400 | 4 | 27·10$^3$ |

Cell size was chosen to be 4 nm, which is lower than calculated exchange length of Co (5 nm). Diameter of the FM phase was changed from 100 nm to 60 nm. AFM thickness was set to 8 nm. Material was divided into 120 different grains with 8 nm size. In each grain the values of exchange stiffness ($A_{ex}$) and anisotropy constants ($K_1$) were randomly varied with maximum deviation of 20 %. Exchange coupling between AFM and FM phases was reduced to 10 %, since in the real material not all of the grains are exchange-coupled [2]. To simulate the textured sample, anisotropy axis in each grain randomly deviated from <001> direction with maximum deviation of 20 °. In the non-textured sample the maximum deviation was set to 60 °.

## Results

Figure S1 shows potential versus time dependence for both current densities. Two areas can be distinguished on the obtained dependencies. Region 1 corresponds to the charge of the electric double layer. The growth of nanorods in pores occurs in region 2, and, as can be seen, the potential remains almost constant. With an increase in the process time, a third region can also be observed on the dependences, at which the potential grows. This region corresponds to the growth of metal not in the pores, but on the membrane surface.

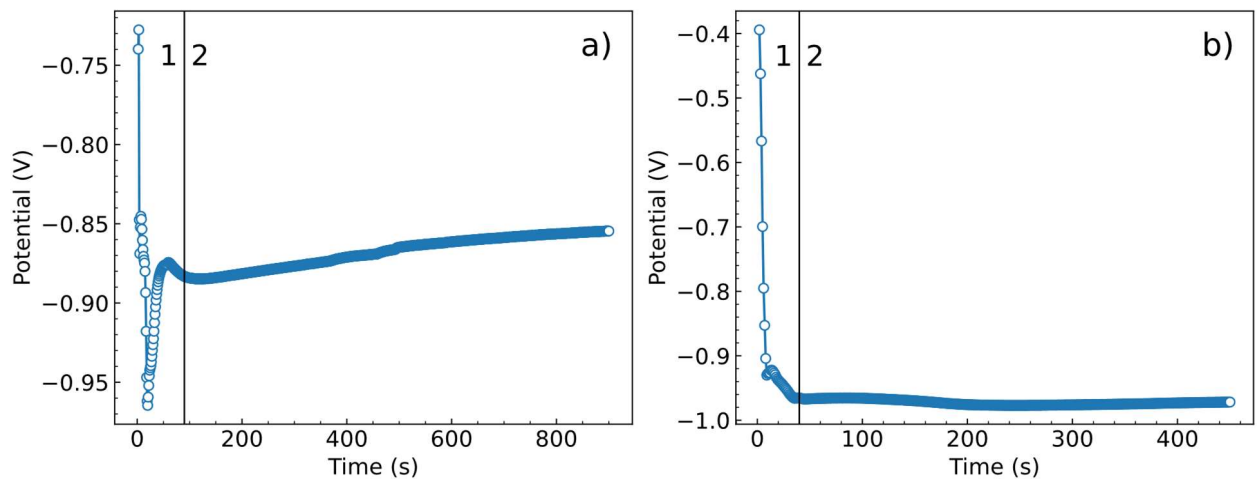

**Figure S1.** Time dependence of the potential during deposition with different current densities. a) 2 mA/cm$^2$. b) 4 mA/cm$^2$.

Figure S2 shows isothermal remanent magnetization (IRM) and DC demagnetization remanence (DCD) curves for samples obtained with different current densities. Prior to measurements the samples were demagnetized by subsequent application of positive and negative fields.

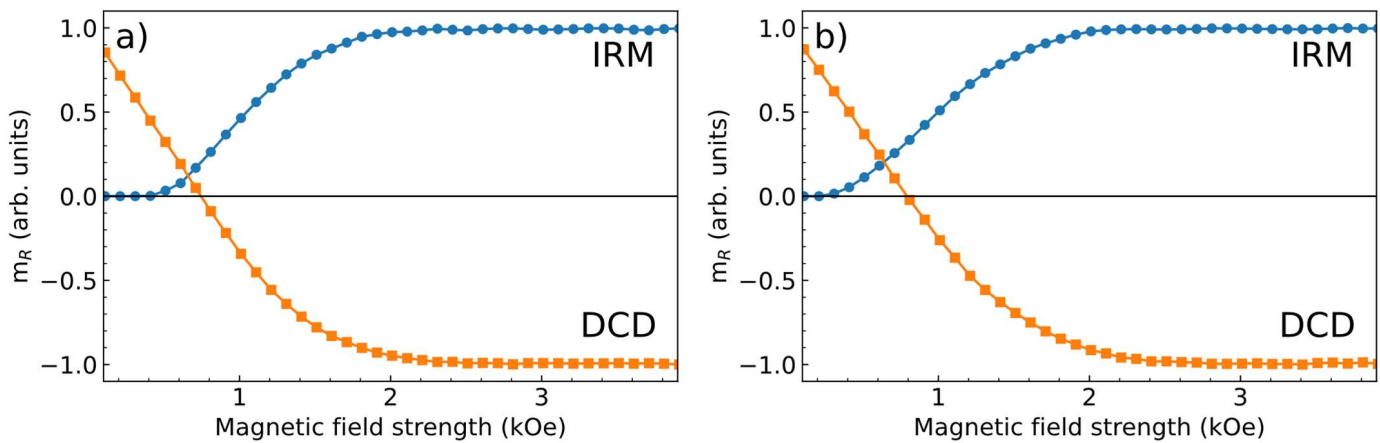

**Figure S2.** Isothermal remanent magnetization (IRM) and DC demagnetization remanence (DCD) curves for samples. a) textured sample. b) non-textured sample.

Figure S3 shows hysteresis loops that were obtained while measuring FORC. These loops were used as an input for FORC calculations in doFORC software.

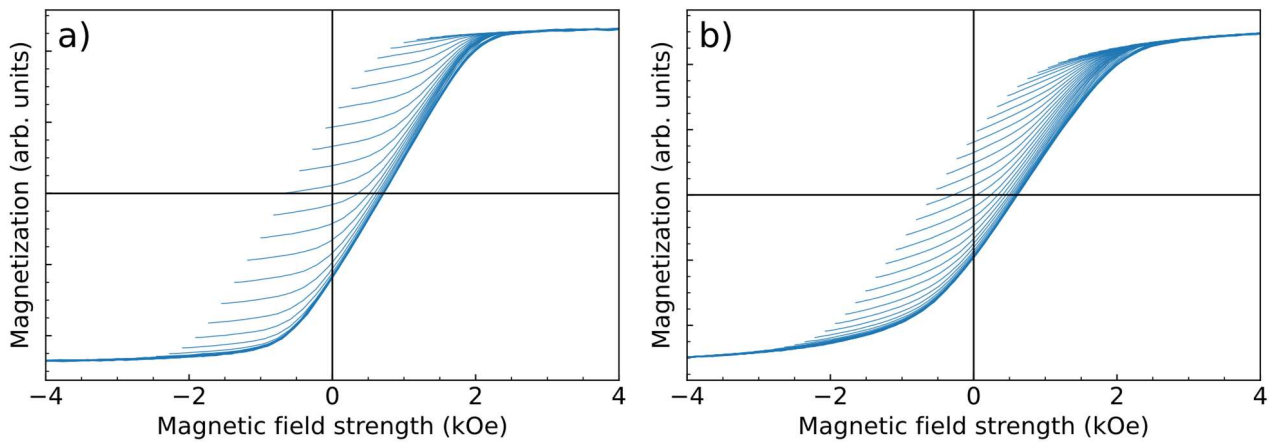

**Figure S3.** Hysteresis loops obtained while measuring FORC for samples. a) textured sample. b) non-textured sample.

Figure S4 shows Transmission Electron Microscopy (TEM) microphotographs of the textured sample after the synthesis. It is seen that the oxide layer is about 18 nm-thick. Electron diffraction pattern shows that the sample has $Co_3O_4$ phase.

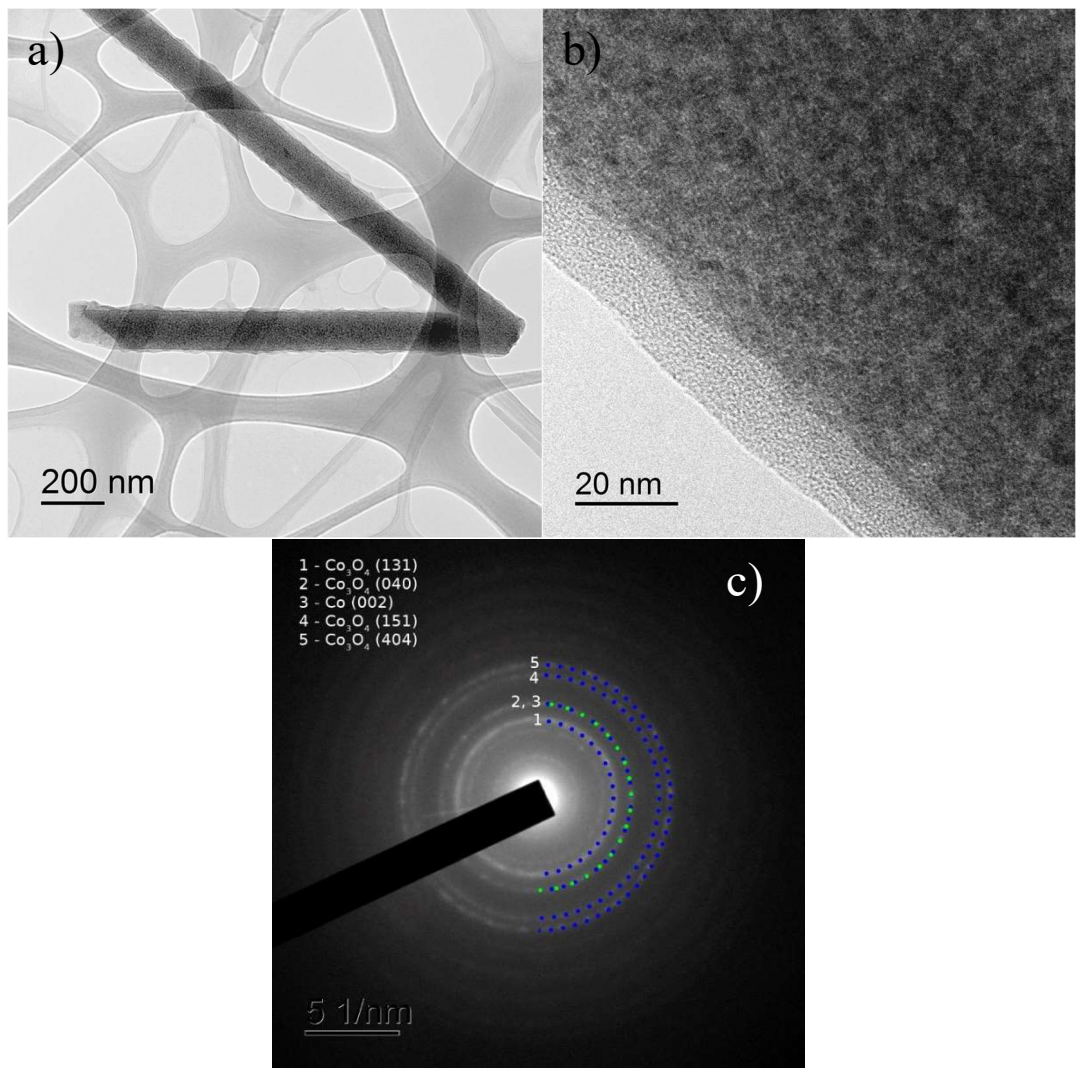

**Figure S4.** Transmission electron microscopy (TEM) photographs of the sample deposited with 2 mA/cm$^2$ current density. a-b). Photos. c) Electron diffraction pattern of area b)

Figure S5 shows Low temperature hysteresis loops of the samples deposited with different current density. The hysteresis loops were taken at 10 K in both ZFC and FC (9 T) mode. ZFC and FC hysteresis loops for both samples were coinciding, meaning that there is no exchange bias between FM and AFM phases.

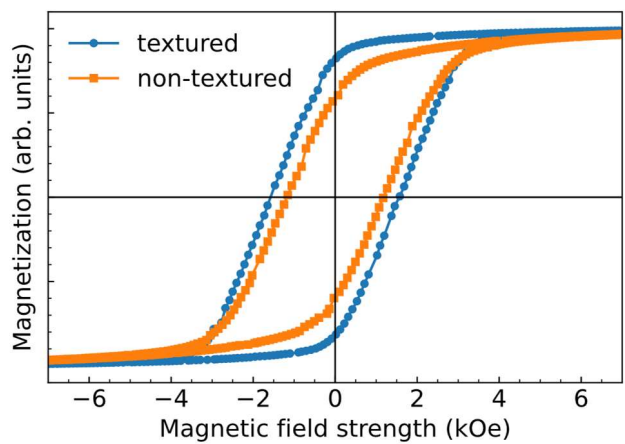

**Figure S5.** Low temperature hysteresis loops of the samples deposited with different current density.

Figure S6 shows Transmission Electron Microscopy (TEM) microphotographs of the sample after the additional oxidation. It is seen that thickness of the oxide layer has increased. Electron diffraction pattern shows that the sample has $Co_3O_4$ phase.

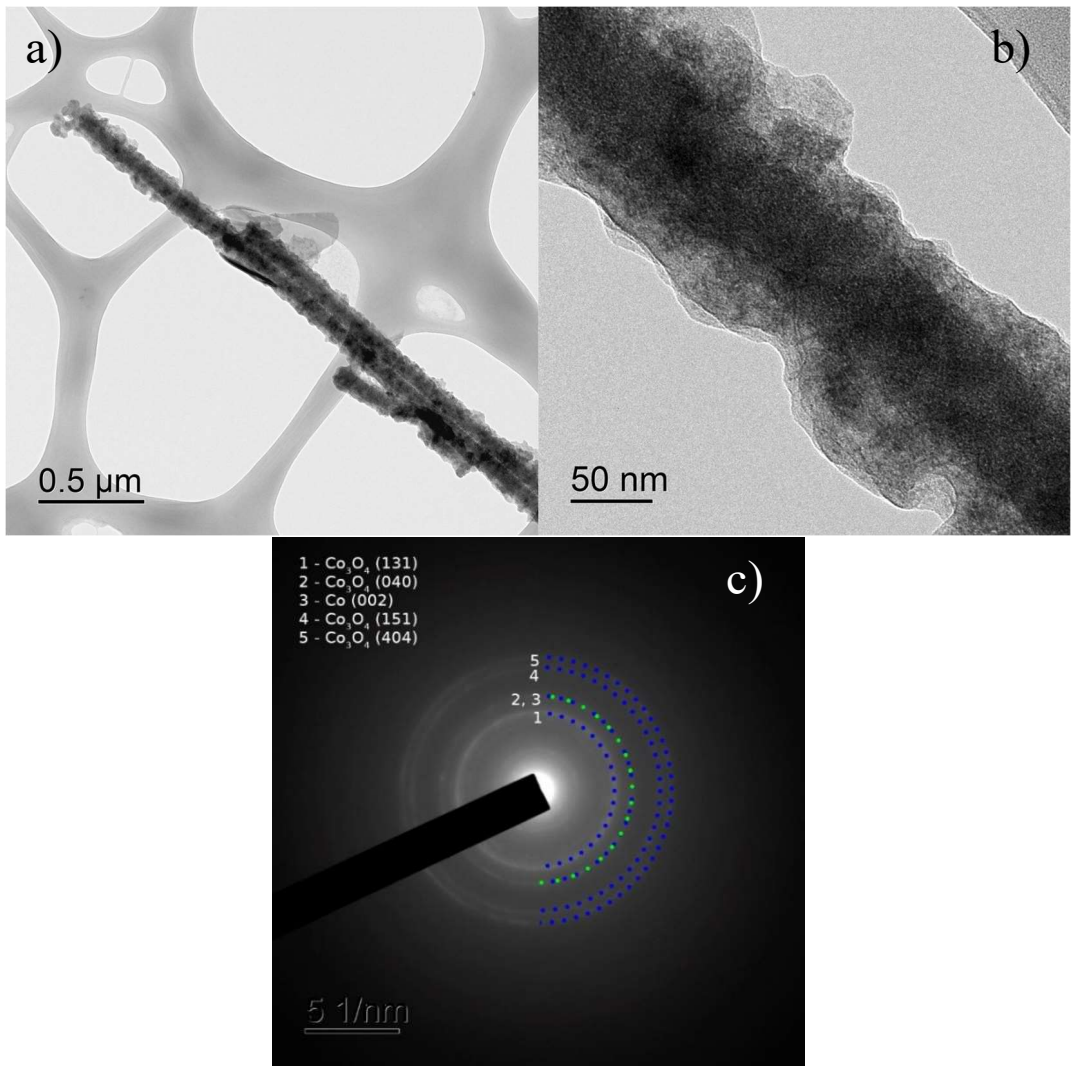

**Figure S6.** Transmission electron microscopy (TEM) photographs of the additionally oxidized sample deposited with 2 mA/cm² current density. a). Photos. b) Electron diffraction pattern of area a)

Figure S7 shows hysteresis loops calculated from micromagnetic simulations in Mumax3.

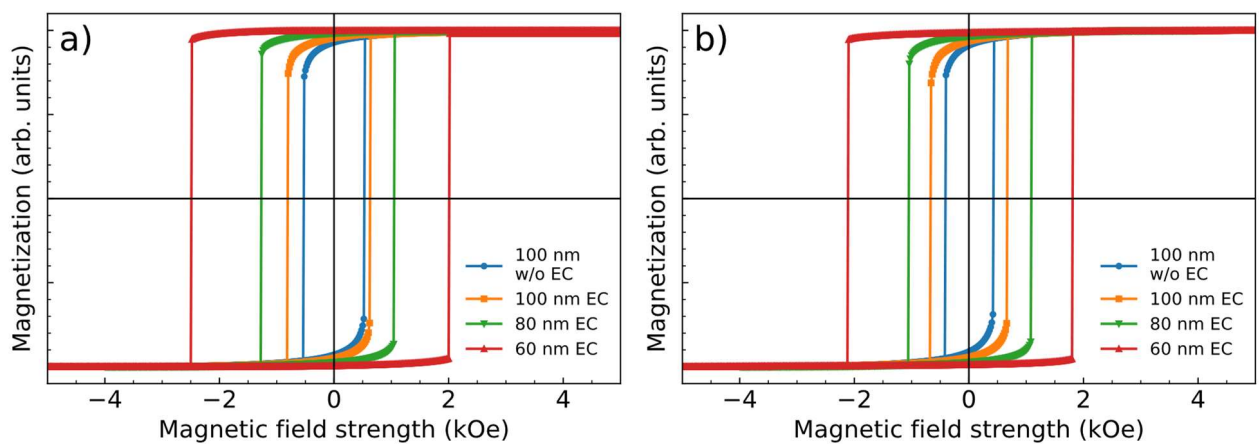

**Figure S7.** Hysteresis loops calculated from micromagnetic simulations in Mumax3. a). Textured nanorod. b) Non-textured nanorod.